# Transport properties of disordered 2D complex plasma crystal

E G Kostadinova, [1] F Guyton, [2] A Cameron, [3] K Busse, [1] C Liaw, [4] L S Matthews, [1] and T W Hyde [1]

[1] Center for Astrophysics, Space Physics & Engineering Research and Department of Physics, One Bear Place 97310, Baylor University, Waco, TX, 76706, USA
[2] Department of Physics, Applied Physics, and Astronomy, Rensselaer Polytechnic Institute, 110 8th St, Troy, NY 12180, USA
[3] Department of Physics, Brigham Young University, Provo, UT 84602, USA
[4] Department of Mathematical Sciences, University of Delaware, 311 Ewing Hall, Newark, DE 19716, USA

E-mail: Eva_Kostadinova@baylor.edu, guytof@rpi.edu, adamc9900@gmail.com, Kyle_Busse@baylor.edu, liaw@udel.edu, Lorin_Matthews@baylor.edu, Truell_Hyde@baylor.edu

**Summary**

In this work, we investigate numerically the transport properties of a 2D complex plasma crystal using diffusion of coplanar dust lattice waves. In the limit where the Hamiltonian interactions can be decoupled from the non-Hamiltonian effects, we identify two distinct types of wave transport: Anderson-type delocalization and long-distance excitation. We use a recently-developed spectral approach to evaluate the contribution of the Anderson problem and compare it to the results of the simulation. The benefit of our approach to transport problems is twofold. First, we employ a highly tunable macroscopic hexagonal crystal, which exhibits many-body interactions and allows for the investigation of transport properties at the kinetic level. Second, the analysis of the transport problem in 2D is provided using an innovative spectral approach, which avoids the use of scaling and boundary conditions. The comparison between the analytically predicted and numerically observed wave dynamics allows for the study of important characteristics of this open system. In our simulations, we observe long-distance lattice excitation, which occurs around lattice defects even when the initial perturbation does not spread from the center to the exterior of the crystal. In the decoupled Hamiltonian regime, this many-body effect can be contributed to the dust lattice interaction with the plasma environment.

# 1. Introduction

Since the discovery of the electric charge, the question of conductivity in various materials has been an active field of research for mathematicians, physicists, and engineers. The comprehensive treatment of a realistic transport problem relies on the subtle interplay between static and dynamical effects. A perfectly ordered crystal at zero temperature acts like a conductor for a traveling electron. Once a critical number of lattice imperfections is present, the wave-particle experiences a crossover from an extended to a localized state, corresponding to a metal-to-insulator transition (MIT) in the medium. In the zero-temperature limit, the interaction between a propagating wavefunction and lattice defects is often described by the class of Anderson-type problems, which includes (but is not limited to) the Anderson localization, quantum percolation, and binary alloy models. In general, Anderson-type Hamiltonians account for transport through short-range interactions. However, in the important case of charged defects (called Coulomb impurities) the analysis should be extended to include long-distance effects, such as collective behavior, magnetic phenomena, and Coulomb blockage. If one further assumes that all lattice points are charged, the system becomes strongly coupled. In this case, the physics of strong correlations in the presence of impurities can be investigated with the help of dynamical mean-field theory and geometric averaging over disorder. Finally, the study of such non-ideal crystals should account for both finite temperature and interactions with the environment, which are characteristic of any open system.

In this paper, we investigate numerically various aspects of the above-mentioned phenomena using a 2D complex plasma crystal. Complex (dusty) plasma consists of charged micron-sized particles (commonly referred to as dust) suspended in a weakly ionized gas [1]–[4]. Dusty plasmas exhibit the collective behavior (including structure formation, self-organization, phase transitions, waves, and instabilities) characteristic of many physical systems. A variety of processes in complex plasma crystals have already been shown to be analogous to those found in other strongly correlated Yukawa systems (for instance, Wigner crystallization in 2D, quantum dot excitation, and excitonic condensation in symmetric electron-hole bilayers [5]–[9]). In this paper, the dust crystal is used as an ideal 'toy' model environment for the study of fundamental problems in transport theory. Key advantages of this approach are: (i) tunability of the crystal disorder over a wide range of system parameters, (ii) macroscopically observable system dynamics, and (iii) the presence of both static and dynamical effects.

Specifically, this study considers the diffusion of coplanar lattice waves in a 2D disordered dust crystal for the regime where Hamiltonian interactions can be decoupled from non-Hamiltonian effects. Here, the Hamiltonian part is represented by the wave interaction with hot-solid and topological defects in a system, while the non-Hamiltonian part arises as a long distance many-body effect, resulting from the dust lattice contact with the plasma environment. In our simulations, we introduce a coplanar wave excitation in the bulk of the crystal and record its space-time evolution. The data is then analyzed using a newly-developed spectral approach, which can determine whether the wave reaches the exterior of the crystal or becomes localized due to spatial defects. The focus of this work is the case where long-distance lattice excitations are observed even when the spectral method does not indicate delocalization of the initial perturbation. In the decoupled Hamiltonian regime, such long-distance excitations can be contributed to the interaction with the plasma gas.

## 2. Theoretical background

A common approach to the dynamics of a non-Hamiltonian problem is to assume that the system of interest is small and coupled to a much larger (but finite) volume, which acts like a thermal bath [10]. In this approximation, the time evolution is governed by a unitary transformation generated by a global Hamiltonian of the form

$$H = H_S + H_B + H_{SB} \tag{1}$$

where $H_S$, $H_B$, and $H_{SB}$ are the Hamiltonians of the system, the bath, and the system-bath interaction, respectively. We introduce three important simplifications to this problem: (i) weak system-bath coupling, (ii) fast bath dynamics (i.e. $H_B = 0$), and (iii) an initially uncorrelated system and bath (i.e. $H_{SB} = 0$ at $t = 0$). When all three approximations are satisfied, the time evolution of the density matrix for the system state $\rho_S(t)$ is reduced to a differential equation of the form

$$\frac{d}{dt}\rho_S(t) = -\frac{i}{\hbar}[H_S, \rho_S(t)] + \mathcal{L}_D\rho_S(t), \tag{2}$$

where $\mathcal{L}_D$ is the generator of dissipative dynamics. Thus, in this limit, transport in the open system can be decoupled into transport in the closed system (first term of equation (2)) plus dissipation to the environment.

Here we examine a classical analogue of the quantum open system problem using a 2D dust crystal in contact with a plasma gas environment. The internal coupling between pairs of dust grains within the crystal is dominated by the repulsive screened Coulomb force, which is in general much larger than the averaged interaction with the plasma flow. In Earth-based experiments, dust structures are levitated against gravity with the help of a vertical electric field. At the same time, the ion flow from the bulk is known to produce a variety of effects such as wakefield focusing, shadowing forces, and dipole polarization of the grains. In most cases, such effects have not been shown to contribute appreciably to the interparticle potential in the horizontal plane. Thus, for the case of a 2D dusty plasma crystal, the in-plane system-bath coupling can be considered weak (when compared with the intergrain potential), which satisfies approximation (i). The requirement of fast bath dynamics given in (ii) is easily met in our simulations, since the time scales of the dust lattice waves are considerably larger than the frequencies of plasma oscillations. Finally, the assumption in (iii) can be (approximately) achieved in both numerical and experimental setups if the 2D crystal is in effective equilibrium at $t = 0$. Here, we define effective equilibrium with the environment as the state in which the thermal velocity fluctuations of the dust grains are much smaller than the propagation velocity of the dust lattice wave.

With the above assumptions satisfied, we argue that transport phenomena in our simulations can be decoupled into two distinct contributions: a Hamiltonian interaction (modeled by an Anderson-type problem) and a non-Hamiltonian effect (resulting from the time-dependent interaction with the environment). In this analysis, the spectral theory will be employed to determine the contribution from the Anderson-type problem. The following section provides a brief overview of the logic behind the spectral approach to transport in 2D systems.

## 3. Spectral approach to the Anderson-type problem

We now outline the setup of the Anderson localization problem and the basic logic of the spectral method. Detailed proofs and physical interpretation of the model can be found in [11]–[13].

Consider the static lattice, zero temperature transport problem, where a wave propagates through a disordered medium. The 2D dust crystal employed in the present numerical simulations exhibits hexagonal symmetry. Thus, here we are interested in the 2D separable Hilbert space $\mathcal{H} = l^2(\Gamma^2)$ of square summable sequences on the hexagonal lattice $\Gamma^2$. (Note that the same spectral approach can be generalized to any dimension and lattice geometry.) The appropriate Hamiltonian for this system is given by the discrete random Schrödinger operator of the form

$$H_\epsilon = -\Delta + \sum_{i \in \Gamma^2} \epsilon_i v_i \langle v_i |, \tag{3}$$

where $\Delta$ is the discrete Laplacian in 2D, $\{v_i\}_{i \in \Gamma^2}$ are the integer lattice site position vectors of the Hilbert space $\mathcal{H}$, and $\{\epsilon_i\}_{i \in \Gamma^2}$ is a set of random variables on a probability space $(\Omega, P)$. The random variables $\epsilon_i$ are independent identically distributed (i.i.d.) in the interval $[-W/2, W/2]$ with probability density $\chi$. The discrete random Schrödinger operator in Equation (3) models a 2D lattice of atoms located at the integer points of a hexagonal lattice $\Gamma^2$. The perturbative part of the operator assigns a random amount of energy $\epsilon_i$ from the interval $[-W/2, W/2]$ according to the prescribed probability distribution $\chi$. Thus, the amount of disorder in the system can be varied by controlling the magnitude of $W$.

For a fixed value of the disorder $W$, the question of interest is whether the energy spectrum of the given Hamiltonian consists of localized or extended states. The energy spectrum can be obtained by diagonalizing the corresponding Hamiltonian operator. The 2D discrete random Schrödinger operator in (3) can be diagonalized with the help of the Spectral Theorem, i.e. through spectral decomposition of the Hilbert space on which the Hamiltonian acts. It can be shown [14] that cyclicity of all vectors in $l^2(\Gamma^2)$ is related to the singular part of the spectrum (i.e. localized states) and that non-cyclicity of any individual vector corresponds to the existence of an absolutely continuous part of the spectrum (i.e. extended states).

<u>Cyclicity</u>: A Hamiltonian $H_\epsilon$ on a Hilbert space $\mathcal{H}$ is said to have a *cyclic* vector $v_0$ if the span of the vectors $\{v_0, H_\epsilon v_0, (H_\epsilon)^2 v_0, \dots\}$ is dense in $\mathcal{H}$.

<u>Spectral Theorem</u>: When $H_\epsilon$ is self-adjoint and cyclic, a unitary operator $U$ exists so that

$$H_\epsilon = U^{-1} M_\xi U, \tag{4}$$

where $M_\xi f(\xi) = \xi f(\xi)$ is the multiplication by the independent variable on another square-integrable Hilbert space $L^2(\mu)$. The new space $L^2(\mu)$ stands for the square-integrable functions with respect to $\mu$ and allows decomposition of the spectral measure $\mu$ into an absolutely continuous part and a singular part

$$\mu = \mu_{ac} + \mu_{sing}. \tag{5}$$

The space $L^2(\mu)$ itself is decomposed into two orthogonal Hilbert spaces $L^2(\mu_{ac})$ and $L^2(\mu_{sing})$. The Hamiltonian has a part $(H_\epsilon)_{ac}$ that comes from $L^2(\mu_{ac})$ and a part $(H_\epsilon)_{sing}$ that corresponds to $L^2(\mu_{sing})$.

<u>Theorem</u> [15]: For any vector $v_0$ in the lattice space, $v_0$ is cyclic for the singular part $(H_\epsilon)_{sing}$ with probability 1.

Theorem [11]: If one shows that $v_0$ is *not* cyclic for $H_\epsilon$ with non-zero probability, then almost surely[1]

$$(H_\epsilon)_{sing} \neq H_\epsilon, \qquad (6)$$

which indicates the existence of extended states.

Spectral method: For a given realization of the disorder $W$ in the system:

(i) Fix a random vector, say $v_0$, in the $2D$ space and generate the sequence $\{v_0, H_\epsilon v_0, (H_\epsilon)^2 v_0, \cdots, (H_\epsilon)^n v_0\}$ where $n \in \{0,1,2,\ldots\}$ is the number of iterations of $H_\epsilon$ and is used as a timestep.

(ii) Apply a Gram-Schmidt orthogonalization process (without normalization) to the members of the sequence and denote the new sequence $\{m_0, m_1, m_2, \ldots, m_n\}$.

(iii) Calculate the distance from *another* vector in the lattice space, say $v_1$, to the $n$ dimensional orthogonal subspace $\{m_0, m_1, m_2, \ldots, m_n\}$, given by

$$D^n_{\epsilon,W} = \sqrt{1 - \sum_{k=0}^{n} \frac{\langle m_k | v_1 \rangle^2}{\|m_k\|_2^2}}, \qquad (7)$$

where $\|\cdot\|_2$ is the Euclidean norm and $\langle \cdot | \cdot \rangle$ is the inner product in the space. It can be shown [11], [12] that extended states exist with probability 1 if

$$\lim_{n \to \infty} D^n_{\epsilon,W} > 0 \qquad (8)$$

is true with a nonzero probability.

## 4. Numerical simulations of dust particle dynamics

To generate the complex plasma crystal, we employed a self-consistent $N$-body code Box_Tree, which provides a user-specified coordinate system, dust particle size, charge, and density, Debye length, external potential, and interparticle forces. This code has been used extensively to model the dynamics of charged dust in astrophysical environments [16]–[20] and in a GEC RF reference cell in Earth-based experiments [21]–[23]. Box_Tree also allows for control over confining potentials in the radial and vertical directions, gravity, ion drag, neutral gas drag, and thermophoretic forces

### 4.1. Dust crystal formation and defect types

In this study, we consider a two-dimensional crystal consisting of $10^4$ identical spherical grains suspended in weakly ionized plasma gas. The dust structure was obtained using system parameters that correspond to experimentally achievable conditions (see Table 1). To ensure that the examined unperturbed crystal is approximately two-dimensional, we require that at the dust grains are in effective equilibrium (i.e. have small average thermal velocities in each direction) and that the vertical extent of the structures is smaller than the average interparticle separation. Starting from initial random positions, the particles were allowed to approach equilibrium by including the effects of a drag force.

---

[1] Note that in probability theory an event happens almost surely if it happens with probability 1. In this paper, we use the two phrases interchangeably.

Figure 1 shows the spatial extent of the dusty structure in the horizontal direction. The observed radial symmetry is a result of the confinement force, applied in the horizontal direction (column five in Table 1). The crystal is levitated at a vertical position of ≈ 5 mm and has average interparticle spacing of ≈ 300 μm, which agrees with experimentally obtained values [4]. The vertical spread of the crystal is less than the particle radius of $r = 5$ μm. Thus, we conclude that the numerically generated complex plasma structure in our simulations is two-dimensional.

To evaluate the amount of lattice defect in the dusty crystal, we employ a crystallinity code, which calculates the true number of nearest neighbors for each dust grain. This information is then used to determine the defect fraction $D$ (defined as the ratio of particles with nearest neighbors different than 6) and the complex bond-order parameter

$$G_6(i) = \frac{1}{6} \sum_{l=1}^{NN_i} e^{i6\Theta_i(l)}, \tag{9}$$

where, $NN_i$ is the number of nearest neighbors of the $i^{\text{th}}$ particle and $\Theta_i(l)$ is the angle of the $l^{\text{th}}$ nearest-neighbor bond measured with respect to the $X$-axis. The determination of the bond order in the crystallinity code relies on the Delaunay triangulation algorithm. Thus, when the code is applied to dust fluid structures with badly defined primitive cells, the Delaunay triangulation function returns an error due to 'insufficient number of unique points'. In other words, in these structures, the function encounters numerous points laying on the same line, in which case, the triangulation does not exist. In this way, we can distinguish between strongly coupled and weakly coupled realizations of the duty structure.

Table 1: Systems parameters used to generate dust crystal at stable equilibrium

| Size [N] | Mass [kg] | Charge [C] | Dust Radius [m] | Conf. [Hz] [2] | Disorder [%] |
|---|---|---|---|---|---|
| $10^4$ | $8.18 \times 10^{-13}$ | $-2.45 \times 10^{-15}$ | $5 \times 10^{-6}$ | $5.4 \times 10^3$ | 5.3 |

[2] Note that the provided frequency is the frequency of inward radial confinement force, i.e. the confinement in the horizontal plane.

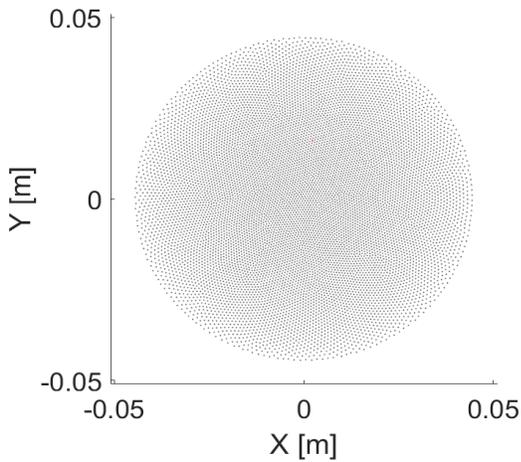

Figure 1: Spatial extent of the numerically generated dust crystal in the horizontal direction.

The crystallinity code was also used to distinguish between the two main variations of spatial disorder (as defined by Thouless [24]): the hot solid type, and the topological disorder. In the static lattice approximation, the hot solid type disorder occurs when atoms are shifted from their regular position in the periodic lattice due to mechanical defect (static positional disorder) or due to the presence of another particle species (substitutional disorder). The examined hot solid type disorder in these simulations is mechanical. Figure 2a shows positions of various mechanical defects within the crystal, where in this case the definition of mechanical disorder coincides with the defect fraction $D$. For the lattice used in this study, we found that $D = 5.3\%$.

In the case of a topologically disordered system, the long-range symmetry in the atomic distribution is completely broken, i.e. uniform periodicity cannot be assumed throughout the lattice. Topological disorder can be represented by a system whose domains exhibit various orientation with respect to the $X$-axis (ranging from $-\pi$ to $\pi$ in radians). Figure 2b shows the formation of characteristic domains throughout the dust structure. In areas where the cell orientation is the same, the color is uniform, whereas areas with changing cell orientation are characterized by color gradient.

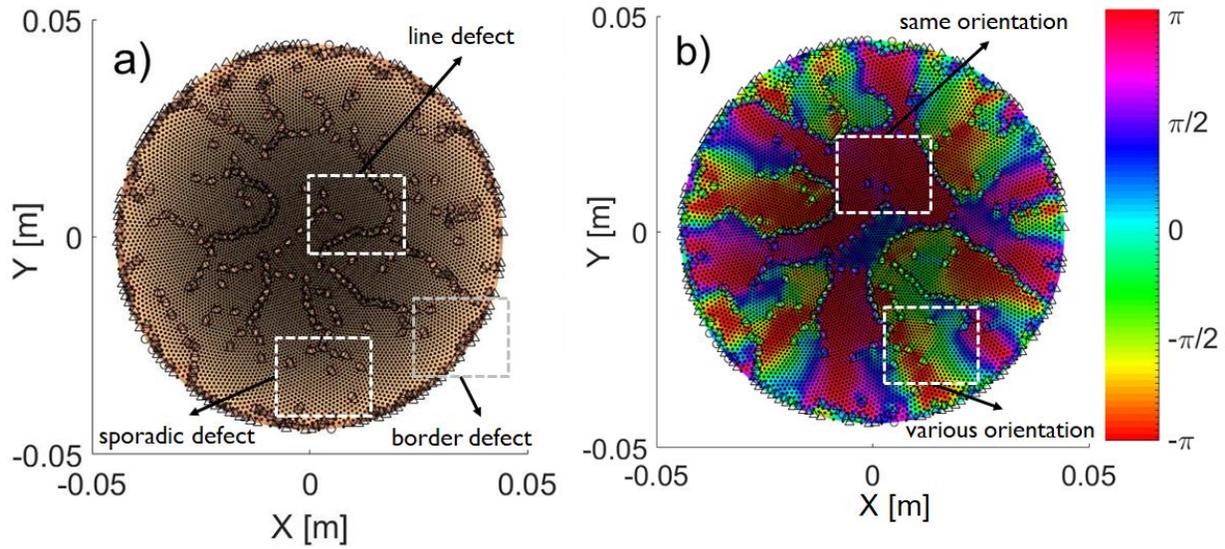

Figure 2: Lattice disorder in the form of mechanical hot solid type defect a) and topological defect b). The positions of the dust grains are marked by dots, with triangles marking particles with $NN = 5$ and circles marking particles with $NN = 7$.

### 4.2. Crystal perturbation

After the crystal has reached equilibrium, the drag force is turned off and a lattice perturbation is induced by an in-plane Gaussian kick of a single particle. Variation of the kick strength, duration, and direction with respect to the X-axis allow for control over the frequency of the propagating lattice wave.

The lattice perturbation in our simulations is induced by an in-plane Gaussian kick of a single particle. Variation of the kick strength, duration, and direction with respect to the $X$-axis allow for control over the frequency of the propagating lattice wave. The magnitude of the perturbation can also be modified using the kick strength or the number of particles initially perturbed. The simulation additionally allows for wave excitation in various areas of the crystal, which can be used to study transport within a specific domain and interaction with the boundary. This paper shows two perturbations of different strength, which were induced in the center of the crystal. In each case, we generated a Gaussian kick at an angle of 0.1336 rad (with respect to the $X$-axis) and a kick duration of 0.04 s. The initial acceleration given to the perturbed particle was 0.016 ms$^{-2}$ in case 1 and 0.030 ms$^{-2}$ in case 2. (Note that in all figures the two cases are labeled $16 \equiv 0.016$ ms$^{-2}$ and $30 \equiv 0.030$ ms$^{-2}$.) The total simulation time was 5 s with an output timestep of

≈ 17 μs (i.t. total number of timesteps $n = 300$). Figure 3 shows maps of the particle trajectories up to the final timestep of the simulation. The trajectory maps in 3a and 3b indicate that in both cases particle excitation was not directly proportional to distance from the kicked particle. Specifically, a comparison between figures 2a,b and 3a,b shows that the grains around defects were displaced more than the grains located in regions of higher crystallinity. This result makes sense as (by definition) particles located near mechanical defects will be shifted from their regular position and will therefore occupy additional unstable energy states.

Figures 3c,d show enlarged maps of the region around the perturbed particle. The small motion of particles at a distance ~10 mm from the perturbed particle suggests that the initial wave perturbation did not spread considerably in the 2D plane. However, the excitation was 'felt' by particles far away from the center of the crystal. In figure 4, we present various plots of the total kinetic energy as a function of time and distance from the perturbed particle. The energy plots further suggest that the initial perturbation was damped out at a radial distance of ~1 mm from the center. This indicates that excitation around defects may contribute to long-range interactions. In the following section, we prove this assertion by using the spectral approach.

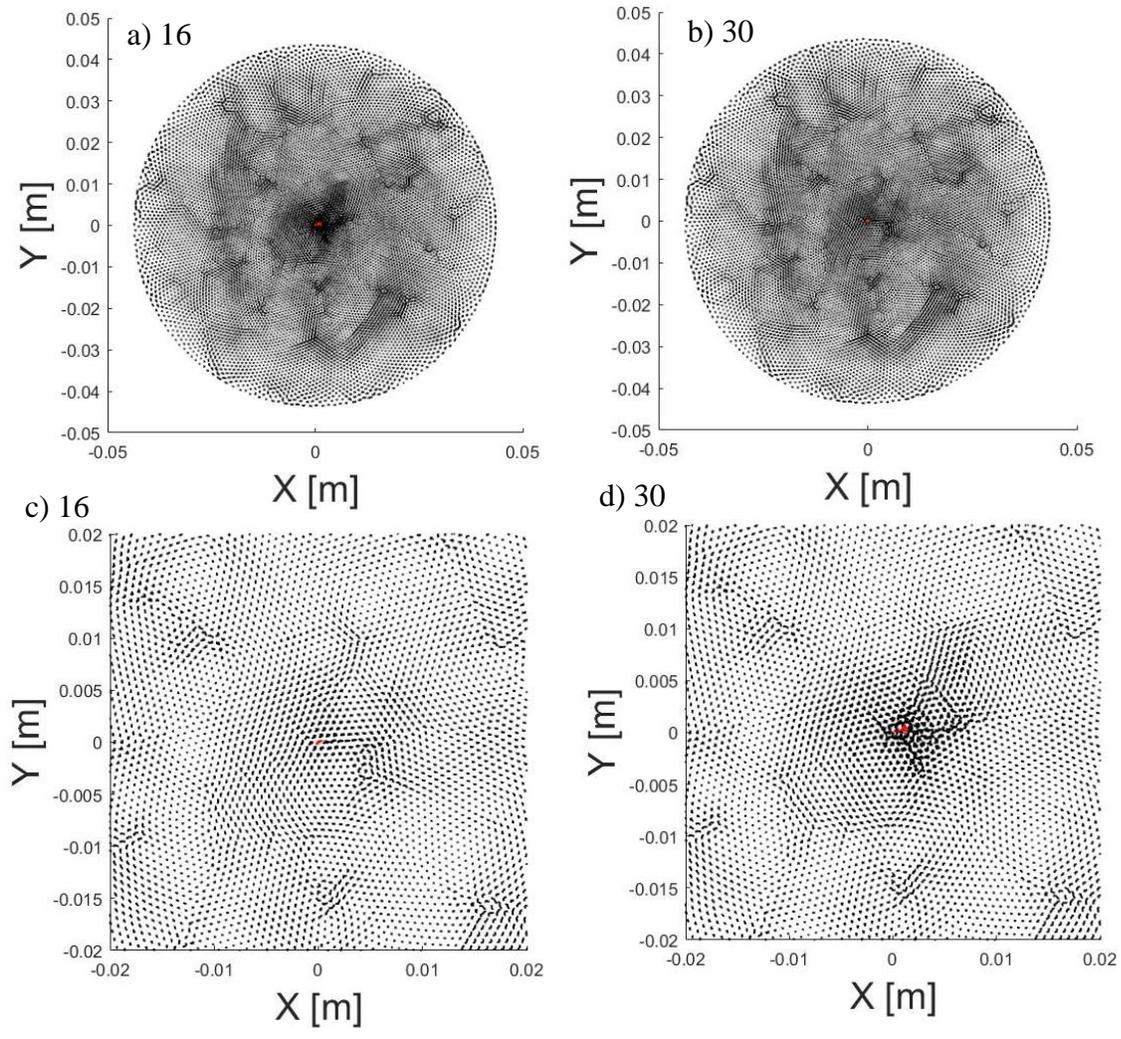

Figure 3: Final movie frames for two different crystal perturbations. Parts a) and b) show the extent of the perturbation throughout the crystal. Parts c) and d) are zoomed-in images of the lattice center, where the perturbation was induced. In all images, the trajectory of the perturbed particle is marked in red.

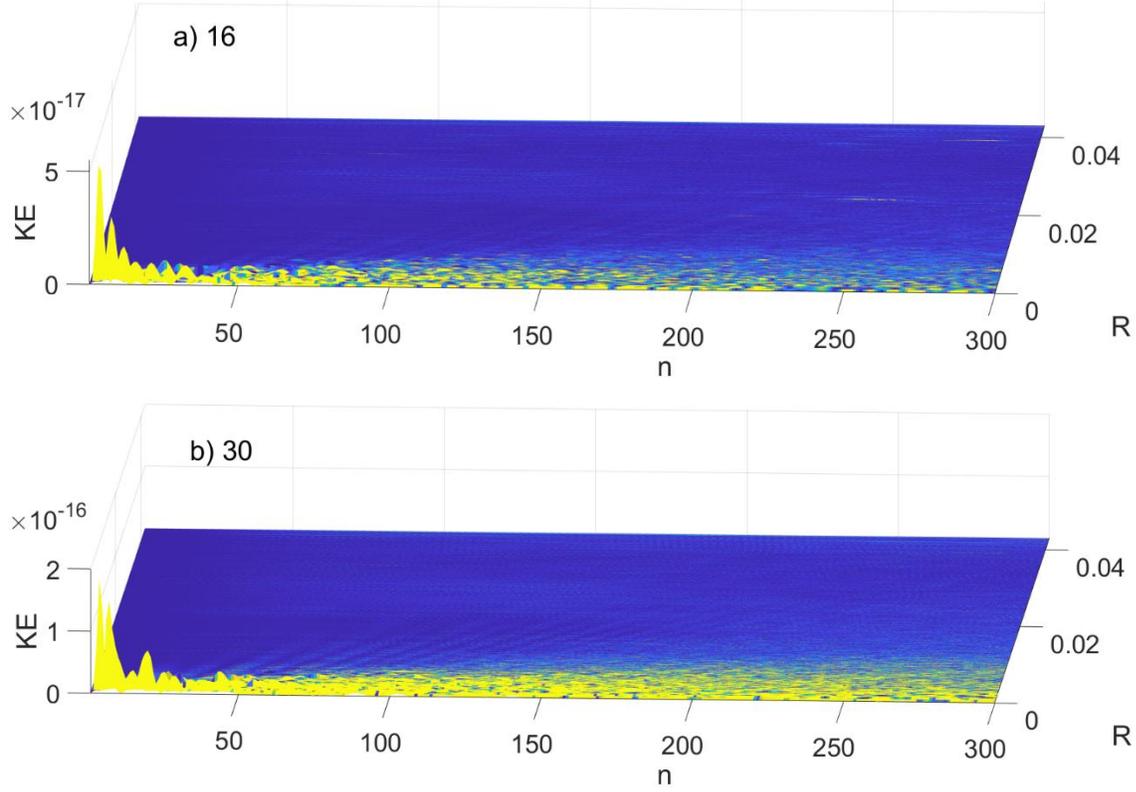

Figure 4: Plots of the total kinetic energy as a function of time and radial distance from the perturbed particle for the two kicks.

### 4.3. Spectral analysis

The sequence $\{v_0, H_\epsilon v_0, (H_\epsilon)^2 v_0, \cdots, (H_\epsilon)^n v_0\}$ from Section 3 represents the dynamical evolution of the perturbed system. In the present case, the vector $v_0$ corresponds to the initial kinetic energy of the crystal right after the Gaussian kick. Each successive term represents the spread of the energy to the nearest neighbors. The numerical equivalent to the operator sequence is given by

$$\{v_0, H_\epsilon v_0, (H_\epsilon)^2 v_0, \cdots, (H_\epsilon)^n v_0\} \rightarrow \{KE_1, KE_2, KE_3, \cdots, KE_n\} \quad (10)$$

where, $KE_n$ is the total kinetic energy of the crystal at timestep $n$. The distance values for the two kicks are calculated using an orthogonalization procedure similar to the one employed in Section 3. Figure 5 shows a plot of the distance values at each iteration $n$. In both cases, the distance value limits to zero with time. According to the criterion given in equation (8), we cannot conclude that the perturbations reached the exterior of the crystal by the nearest-neighbor interaction assumed in the Hamiltonian of equation (3). Thus, we expect that the observed excitations in our simulations do not result from the classical Anderson-type transport. Instead, they can be contributed to a long-distance interaction in the strongly coupled Coulomb system.

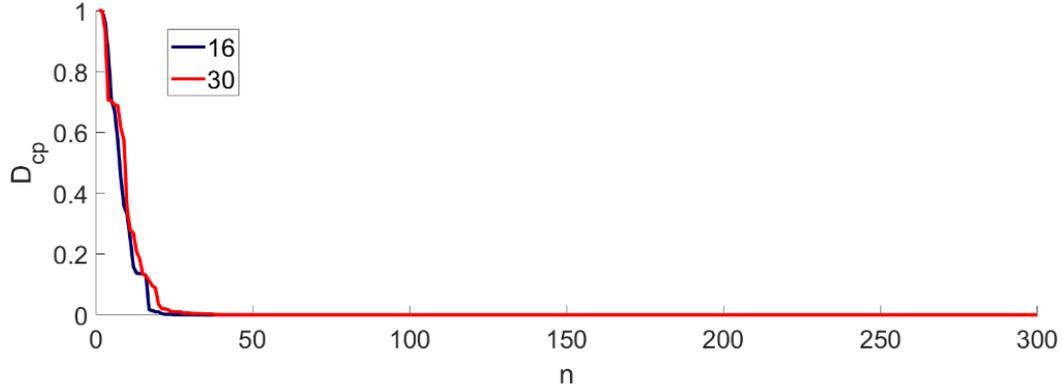
Figure 5: Limiting behavior of the distance values for the two kicks.

## 5. Discussion

The theoretical and numerical analysis presented in this work relies on the decomposition of the open system dynamics into Hamiltonian transport and non-Hamiltonian dissipation to the environment. As mentioned in Section 2, one of the requirements needed for such decomposition is the assumption that at the unperturbed crystal is in effective equilibrium, which ensures that the crystal-plasma interactions are negligible at $t = 0$ (i.e. the velocity fluctuations of the dust grains do not cause significant dissipation to the environment). This (approximate) zero-temperature scenario also allows for the use of an Anderson-type Hamiltonian in the modeling of the closed system dynamics. Thus, for the purposes of the theoretical analysis, we aimed at minimizing the initial velocities of the dust grans in the unperturbed crystal. In the numerical simulation, the average thermal velocity of the particles at equilibrium are $\sim 10^{-8} \text{ms}^{-1}$ in the $X$ and $Y$ directions and $\sim 10^{-16} \text{ms}^{-1}$ in the $Z$ direction indicating that the chosen crystal equilibrium is (effectively) static. These values are several orders of magnitude smaller than the thermal fluctuations usually observed in laboratory conditions ($\approx 10^{-4} \text{ ms}^{-1}$ [4]). However, it has been recognized that thermal effects do not interfere significantly with lattice propagation in cases where the dust grains only move distances smaller than the interparticle separation and have velocities randomly fluctuating in all directions [25]. Thus, we expect that the main results from the presented study should be experimentally observable even in the presence of larger thermal velocities.

## 6. Conclusions and future work

In this paper, we examined numerically the propagation of in-plane dust lattice waves in a two-dimensional disordered complex plasma crystal. The observed crystal dynamics is decomposed into a Hamiltonian transport (in the form of Anderson-type wave delocalization) and non-Hamiltonian interaction (in the form of a long-distance excitation resulting from coupling to the plasma environment). Such decomposition is valid in the case of weak system-bath coupling, fast bath dynamics, and initially uncorrelated system and bath. The contribution from the Anderson-type delocalization is evaluated with the help of an innovative spectral approach. Here we provided two cases where long-distance excitations around lattice defects were observed even though the spectral analysis indicated that the initial perturbation did not spread from the bulk to the exterior of the crystal. Such excitations can be contributed to the dust interaction with the plasma environment.

In our future work, we will extend the numerical simulations to include a wide range of system regimes and types of perturbations. Specifically, we will examine the dependence of long-distance excitations on the defect concentration, cell orientation, and fluctuations of the average dust grain velocities. The numerical results will be compared to laboratory experiments employing laser perturbations induced in a 2D complex plasma crystal. The combined numerical and experimental results will be incorporated into a comprehensive analysis of transport in strongly coupled non-Hamiltonian systems.

## 7. Acknowledgments

This work was supported by the NSF/DMS (grant number 1700204, C D L) and NASA/NSF/DOE (NASA grant number 1571701 and NSF/DOE grant numbers 1414523, 1740203, 1262031, L S M and T W H).